\documentclass[prl,floatfix,twocolumn,showpacs,a4]{revtex4}

\usepackage{graphicx}
\usepackage{epstopdf}
\usepackage[latin1]{inputenc}

\usepackage{graphicx}%
\usepackage{dcolumn}
\usepackage{amsmath}
\usepackage{amssymb}
\usepackage{tabularx}
\usepackage{subfigure}
\begin{document}

\newcommand{\be}{\begin{equation}}
\newcommand{\ee}{\end{equation}}
\newcommand{\bea}{\begin{eqnarray}}
\newcommand{\eea}{\end{eqnarray}}
\newcommand{\pd}{p_{d}}
\newcommand{\pr}{p_{r}}
\newcommand{\pw}{p_{w}}
\newcommand{\dt}{\Delta t}
\newcommand{\wo}{w_0}
\newcommand{\ptri}{p_\Delta}
\newcommand{\timeder}[1]{\frac{\textrm d #1}{\textrm d t}}
\newcommand{\ave}[1]{\langle #1 \rangle}
\newcommand{\neigh}[2]{N_#1 (#2)}
\newcommand{\pto}{P}
\newcommand{\ptoa}{\tilde {\pto}}
\newcommand{\ptrid}{\tilde {\ptri}}
\newcommand{\deltad}{\tilde \delta}
\newcommand{\node}{\nu}
\newcommand{\link}{l}
\newcommand{\rlcc}{R_{LCC}}
\newcommand{\rlck}{R_{k=4}}
\newcommand{\nst}{\ave {n_s}}

\title[Short Title]{Emergence of communities in weighted networks}
\author{Jussi M. Kumpula$^1$}
\email{jkumpula@lce.hut.fi}
\author{Jukka-Pekka Onnela$^{2,3,1}$}
\author{Jari Saram\"{a}ki$^1$}
\author{Kimmo Kaski$^1$}
\author{J\'{a}nos Kert\'{e}sz$^{1,4}$}
\affiliation{%
$^1$Laboratory of Computational Engineering, Helsinki University of Technology, P.O. Box 9203, FIN-02015 HUT, Finland \\
$^{2}$Physics Department, Clarendon Laboratory, Oxford University, Oxford OX1 3PU, U.K.\\
$^{3}$Sa\"\i d Business School, Oxford University, Oxford OX1 1HP, U.K.\\
$^4$Department of Theoretical Physics, Budapest University of Technology and Economics, Budapest, Hungary
}%

\date{\today}
\begin{abstract}
Topology and weights are closely related in weighted complex networks 
and this is reflected in their modular structure. We present a
simple network model  where the weights are generated dynamically and
they shape the developing topology. By tuning a model parameter governing 
the importance of weights,
the resulting networks undergo a gradual structural transition from a
module free topology to one with communities. The model also reproduces
 many  features of large social networks, including the "weak links" property.
\end{abstract}
\pacs{89.75.Hc, 87.16.Ac, 89.65.-s,89.75.Fb,89.75.-k}

\maketitle

Network theory has undergone a remarkable development over the last decade and 
has contributed significantly to our understanding of complex systems, ranging from genetic transcriptions to the Internet and human societies \cite{RefWorks:133,Caldarelli}. 
Many complex networks  are structured in terms of modules, or communities, which are groups of nodes characterized by having 
more internal than external connections between them. Such a mesoscopic network structure is expected to play a 
concrete functional role. Consequently, it is an important problem to understand how the communities emerge during the growth of the 
network. Apart from these issues of topological nature, it is important to realize that many complex networks are weighted, i.e., the interaction between two nodes is  characterized not only by the existence of a link 
but a link with a varying weight assigned to it. There are a number of  examples, like traffic, metabolic or
correlation based networks, which provide ample evidence that the weights have to 
be included in their analysis. In many cases the  weights affect significantly the
properties or function of these networks, e.g.,  disease spreading \cite{VittoriaColizza02142006}, synchronisation dynamics of oscillators \cite{bernardo:2005a},  and motif statistics \cite{onnela:2005}. 
It is natural to expect that weights  have an influence on the formation of communities, which is the very issue of our study.

Earlier, coupled weight-topology dynamics have been used successfully in transport 
networks modeling \cite{RefWorks:131}, which, however, does not lead to community structure.
We show that there are mechanisms, by which weights play a crucial role in community formation. 
While we believe this to be quite 
a general paradigm for community
formation, we have chosen to explore it within the realm of social systems
where large datasets have enabled looking into both the coupling of network
topology and interaction strengths  and properties of
communities \cite{RefWorks:129, RefWorks:52,RefWorks:114}. Understanding how the underlying
microscopic mechanisms translate into mesoscopic communities and macroscopic
social systems is a key problem in its own right and one that is accessible
within the scope of statistical physics.

\begin{figure}[!t]
\includegraphics[width=0.75\linewidth]{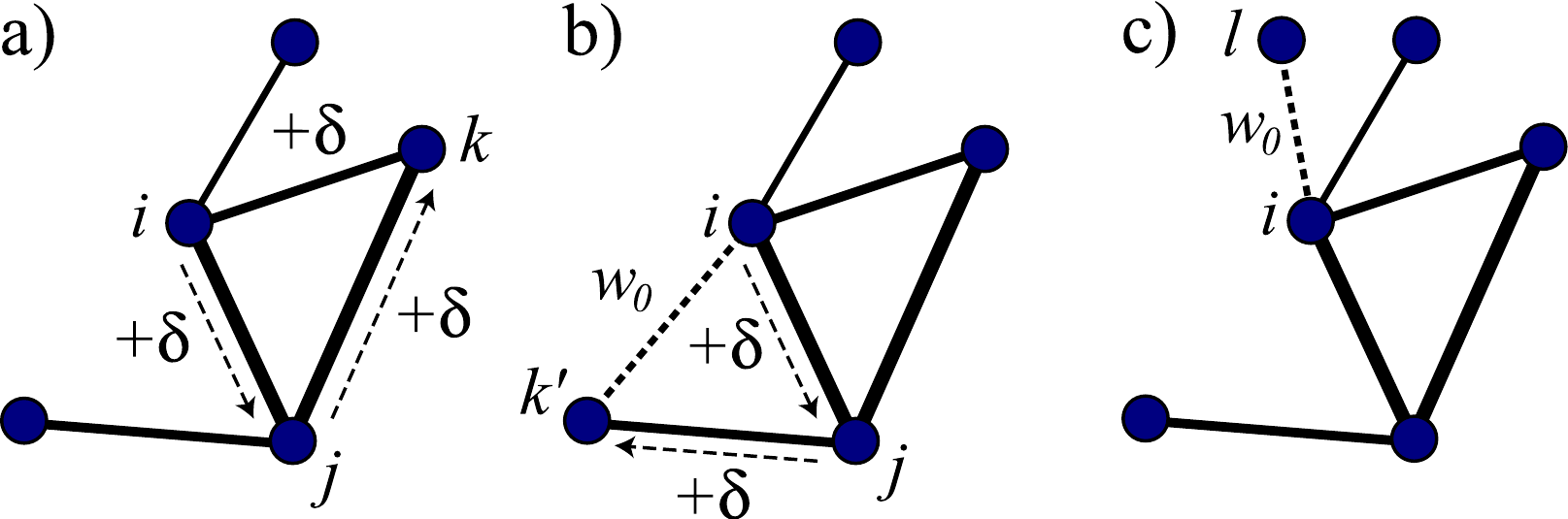}
\caption{The model algorithm. 
(a): a weighted local search starts from $i$ and proceeds first to $j$ and then to 
$k$, which is a neighbor of $i$. 
(b): the local search from $i$ ends to ${k'}$, 
which is not a neighbor of $i$. In this case link $w_{ik'}$ is established with probability $\ptri$.
(c): node $i$ creates a random link to random node $l$ with probability $\pr$.
In cases a) and b) the weights of involved links  are increased by $\delta$.
}
\label{fig:algoritmi}
\end{figure}

Large scale social networks are known to satisfy the weak links hypothesis \cite{RefWorks:128} with the implication that weak links keep the network connected whereas strong links are mostly
associated with communities \footnote{The weak link hypothesis \cite{RefWorks:128} was recently proved on societal-level one-to-one communication networks~\cite{RefWorks:129,RefWorks:130}}. 
 This weight-topology coupling results from the microscopic mechanisms that govern the evolution of social networks.
 Network sociology identifies (a) \emph{cyclic closure} and (b) \emph{focal closure} as the two fundamental mechanisms of tie formation \cite{kossinets:2006}. 
Cyclic closure refers to forming ties with one's network neighbors - "friends of friends". 
Experimental evidence indicates that the probability of cyclic closure decreases roughly exponentially as a function of the geodesic distance \cite{priv:2007}. 
Focal closure, in contrast, refers to forming ties independently of the geodesic distance and is attributed to forming social ties through shared activities (hobbies etc.)  \cite{kossinets:2006}. 
These two mechanisms form the basis of the \emph{topological} rules of our model. 
As for the weights, we have chosen a simple scenario in which new ties are created preferably through strong ties,
every interaction making them even stronger.

\begin{figure}[tb]
	\centering
        \includegraphics[width=0.9\linewidth]{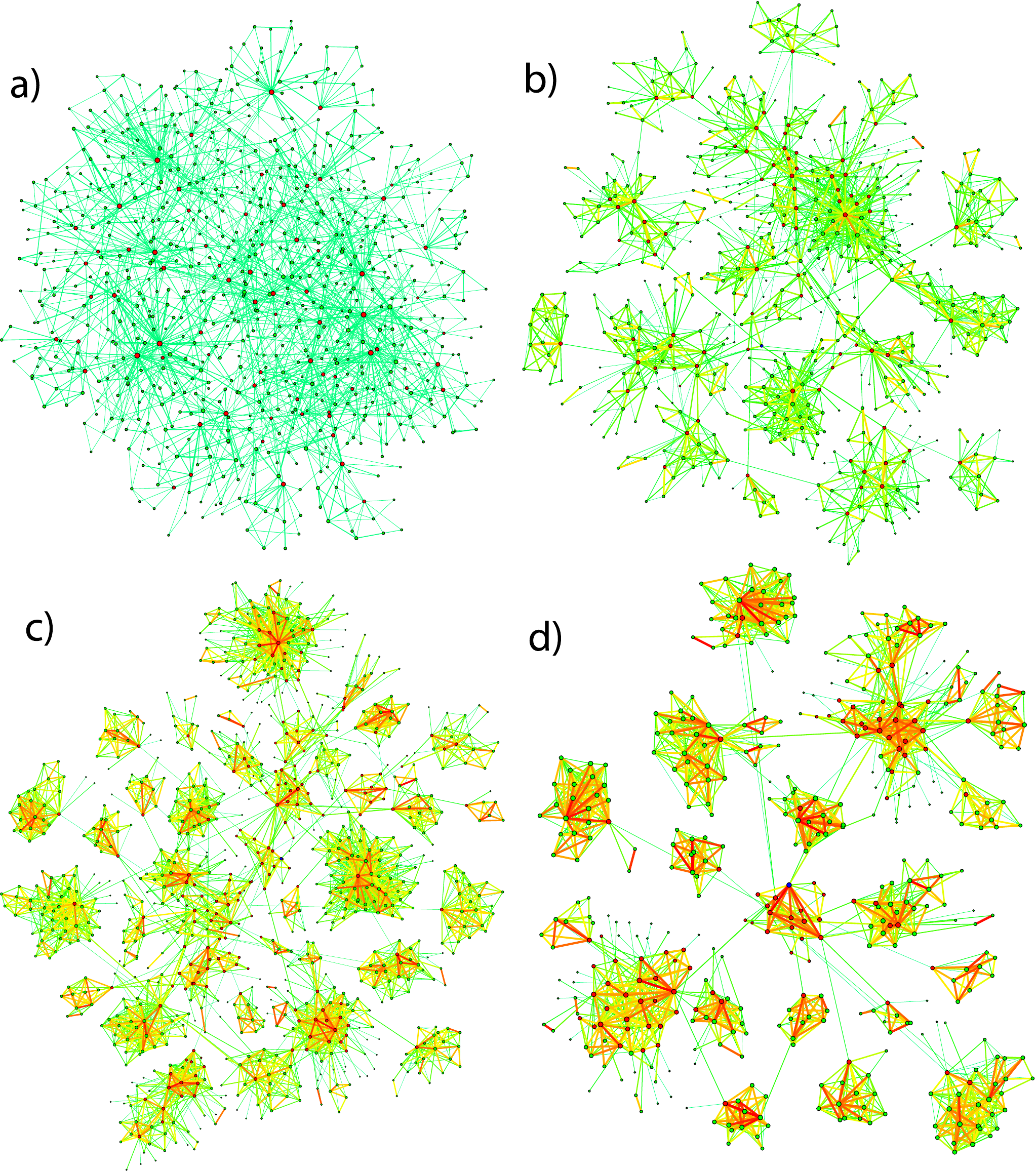}
	\caption{(Color online) Snowball samples \cite{RefWorks:103} of networks with (a) $\delta=0$, (b) $\delta=0.1$, (c) $\delta=0.5$, and (d) $\delta=1$.
	Link colors change from green (weak links) to yellow and red (strong links). }
	\label{fig:yhdistelmakuva2}
\end{figure}

We consider a fixed size network of $N$ nodes,
where links are created in two ways:
First, in time interval $\dt$ each node having at least
one neighbor starts a weighted local search for new acquaintances, Fig. \ref{fig:algoritmi}(a,b).
More specifically, node $i$ chooses one of its neighbors, node $j$,
with probability $w_{ij}/s_i$, where $w_{ij}$ is the weight of the link
connecting $i$ and $j$ and $s_i=\sum_j w_{ij}$ the strength of $i$.
If the chosen node $j$ has other neighbors apart from $i$, it
chooses one of them, say $k$, with probability $w_{jk}/(s_j - w_{ij})$. 
Therefore, the search favors strong links. If there is no link between
$i$ and $k$, it is established with probability
$\ptri \dt$ such that $w_{ik}=w_0$. If the link exists,
its weight is increased by an amount $\delta$. In both cases,
$w_{ij}$ and $w_{jk}$ are also increased by $\delta$.
We call this attachment mechanism \emph{local attachment} (LA), and it corresponds to a special case, triadic closure, of  the
above mentioned cyclic closure mechanism. Second, if a node has no links, or otherwise with probability $\pr \dt$, it creates a link of weight $\wo$ to a random node, Fig. \ref{fig:algoritmi}(c). This mechanism corresponds to establishing a new interaction outside the immediate neighborhood of the chosen node, analoguously to focal closure, and we call it \emph{global attachment} (GA). 
Finally, any node can be removed with probability $\pd \dt$ by the \emph{node deletion} (ND) mechanism, in which case also the adjacent links are removed. 
The removed node gets replaced by a new node, such that the size of the system remains fixed at $N$. 
ND is the only mechanism that decreases the number of links in the model and leads to exponentially distributed lifetime $\tau$ for nodes and $\tau_{w}$ for links, averages of which are 
\bea
\ave{\tau} &=&  (\pd \dt)^{-1} \nonumber \\ 
\ave{\tau_{w}} & = &  (2\pd \dt)^{-1}.
\eea

The model was studied by simulations, which were started from a seed network of $N$ nodes without any links. 
Subsequent changes due to LA and GA mechanisms were updated using parallel update, followed by the ND step.
We set $\dt=1$ and $\wo=1$ without loss of generality.
The time scale is fixed by the death rate $\pd$, which was set to $10^{-3}$ corresponding to $\ave{\tau}=10^3$ time units. 
The random link probability was set to $\pr=5\times10^{-4}$, corresponding to adding on average one random link for each node during average node lifetime. 
The network algorithm was found to reach steady state in $\sim 10-20$ average node lifetimes, after which all measured characteristics were found to be stationary. The following results were obtained  by running the simulations for $25\times 10^3$ time steps, i.e., 25 average node lifetimes. 

The weights enter the model dynamics through parameter $\delta$. In order to compare networks resulting 
from different values of $\delta$, we have chosen to keep the average degree $\ave k \approx 10$ constant. 
Thus, the number of links is roughly equal in all networks and changes in the structure result solely from restructuring of the 
links. Keeping $\ave k$ constant requires adjusting $\ptri$ for each $\delta$, which can be done easily as their dependence
is a smooth monotonous function \footnote{Keeping $\pd$ and $\pr$ constant, the balance of link addition and deletion depends on the rate at which new links are formed in the LA process. Evidently this depends on the LA probability, $\ptri$, but also on the probability to end up at a node not connected to the starting node. Increasing $\delta$ constrains the searches to 'familiar paths', decreasing this probability, and consequently leads to a lower degree.}. 
When $\delta=0$ we obtain unweighted networks that  resemble those obtained by certain older models \cite{RefWorks:84,RefWorks:86} 
without apparent community structure. However, increasing $\delta$ results in denser networks in which communities clearly appear  
as seen in Fig.~\ref{fig:yhdistelmakuva2}. This can be attributed to the effect of $\delta$ on the LA mechanism. 
The higher the value of $\delta$, the more trapped the local searches become, i.e., they repeatedly follow the same links, 
simultaneously increasing the weights of these links and the associated triangles, which in turn 
further enhances the trapping effect.  
Thus, in the transient phase of the model dynamics prior to the stationary state, any emerging triangle starts to rapidly
accumulate weight, acting as a nucleus for community formation.

\begin{figure}[!tb]
	\centering
\includegraphics[width=1.0\linewidth]{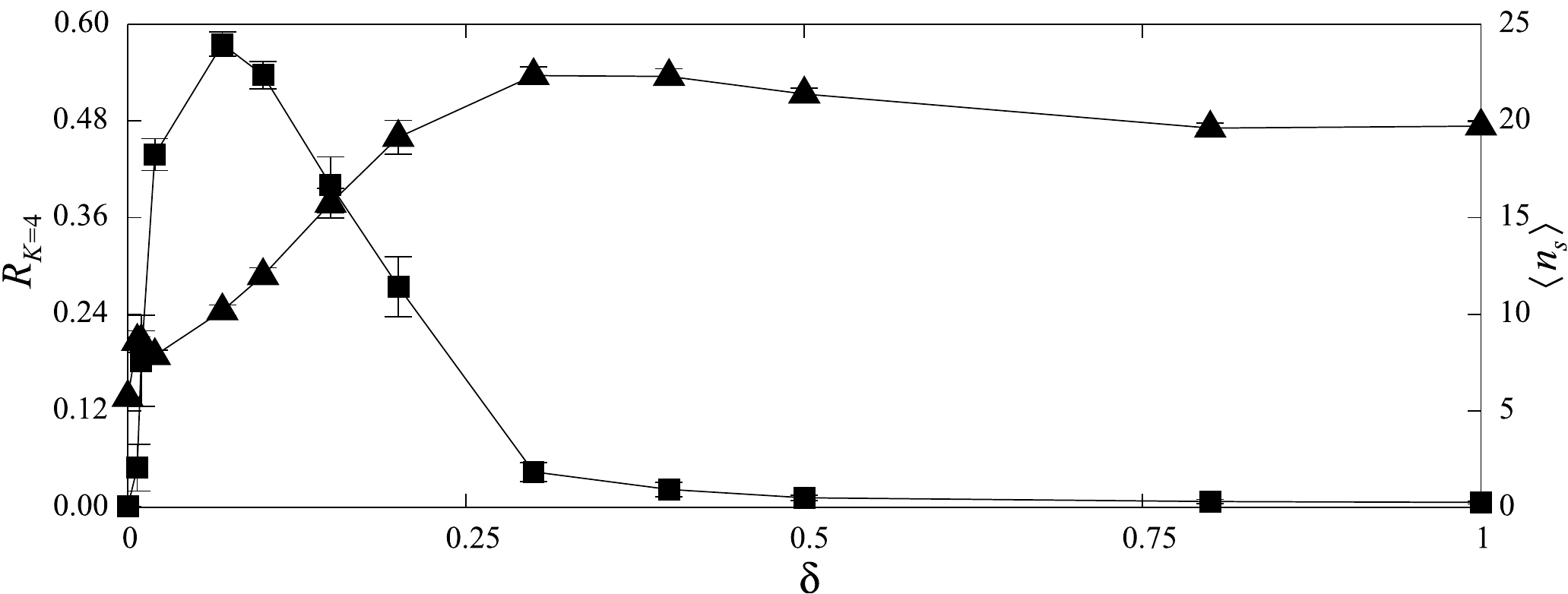}
	\caption{$\rlck$ ($\Box$) and  $\nst$ ($\triangle$)  as a function of $\delta$. 
          Results are averaged over 10 realizations of  $N=5\times 10^4$ networks. 
	Error bars are measured standard deviations.}
	\label{fig:largestComSizeAndMeanSize_constDegHigh_4}
\end{figure}

We consider first the topological structure of the model networks and study the communities
using the $k$-clique percolation method \cite{RefWorks:52,RefWorks:78}. This
avoids the problems 
of modularity-based methods~\cite{RefWorks:46,RefWorks:104,RefWorks:105}, which may not properly resolve 
communities if their size distribution is broad. 
The LA mechanism, which is mainly responsible for introducing new links, generates at least one triangle per added link. Therefore, we focus on 4-cliques, the smallest non-trivial cliques  beyond triangles. Figure~\ref{fig:largestComSizeAndMeanSize_constDegHigh_4} shows the relative largest community size $\rlck$ and average community size excluding largest community $\nst$ for $\delta\in [0,1]$. 
When $\delta=0$ communities are mainly very small, $\nst\sim6$ and the largest ones contain $\sim 50$ nodes. 
Increasing $\delta$ changes the network structure significantly. At first, the network becomes homogeneous in the sense that the 4-cliques percolate through most of the system, but as $\delta$ becomes larger the nodes begin to condensate in tighter communities. 
This can be seen as the increase in $\nst$ and simultaneous decrease in $\rlck$.
When $\delta \gtrsim 0.2$ the network contains communities whose $\nst \sim 20$ while the largest ones consist of several hundred nodes. The probability for communities of size larger than 50 to occur is several orders of magnitude higher for $\delta \gtrsim 0.2$ than for $\delta=0$. Similar results were obtained for $k$-clique percolation with $k=3$ and $k=5$. 
The community  size distribution was found to be stable after $\sim 10$ average node lifetimes. This
can be understood in the large $\delta$ limit by considering the change in the size $N_s$ of community $s$.
When $\delta$ is large the LA process mostly follows the strong within-community links and 
we can assume that community  merging is rare. We can now estimate the change in $N_s$ as
\be
\timeder{N_s} = -\pd N_s + \pd N \frac {N_s} N = 0,
\ee
where the first term on the $r.h.s.$ follows from the fact that each of the $N_S$ nodes in $s$ is
deleted with probability $\pd$, and the second reflects the fact that of the all $\pd N$ deleted nodes,
a fraction of $N_s/N$ will form an initial, random link to community $s$, thus becoming a member of that community
at subsequent time steps by the LA process.
This shows that once the algorithm has reached a state in which most local searches remaining in the initial community is valid, the community size distribution remains constant.
Figures  \ref{fig:yhdistelmakuva2} and \ref{fig:largestComSizeAndMeanSize_constDegHigh_4} indicate that
 $\delta \gtrsim 0.5$ can already be regarded as ``large'', because the communities are tight and increasing $\delta$ does not change them significantly.

\begin{figure}[tb]
	\centering
	\includegraphics[width=1.00\linewidth]{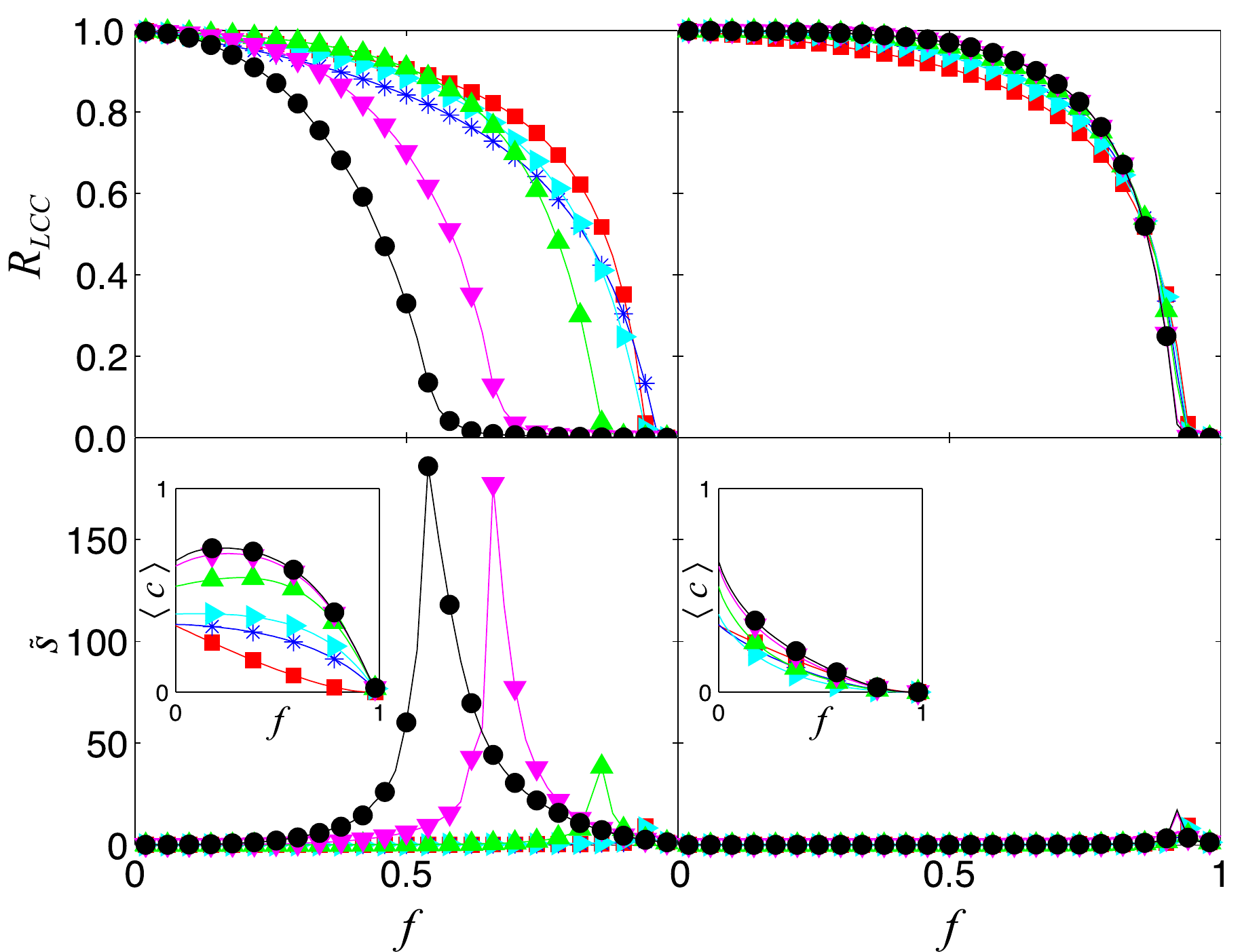}
	\caption{ (Color online) $\rlcc$ and $\tilde s$ for link percolation. Left: weak links removed first, right: strong links removed first.
	Insets: Average clustering. Results are averaged over 10 realizations of $N=5\times 10^4$ networks.
      Values of $\delta$ are 0 ($\Box$), $1\times 10^{-3}$ ($\ast$), $1\times 10^{-2}$ ($\triangleright$), 0.1 ($\triangle$), 0.5 ($\bigtriangledown$), and 1 ($\circ$).}
	\label{fig:GC_and_suskis_constDegHigh}
\end{figure}

Next, we consider the effect of $\delta$ on weight-topology correlations and
study the network structure with link percolation. 
Weak links hypothesis  implies that links within communities are strong whereas links between them are weak. 
Therefore, if the network has such a structure, it should disintegrate faster when links are removed in ascending than in descending order of weight, as observed in \cite{RefWorks:129}. 
Here we remove links from the network in both orders while monitoring the network properties as a function of fraction of removed links $f$, which acts as a control parameter. 
We have measured the relative size of the giant component $\rlcc$ serving as an order parameter, the 'normalized susceptibility' $\tilde s=\sum n_s s^2/N$, 
where $n_s$ is the number of components of size $s$ and the sum is taken over all but the largest component, and the
average clustering coefficient $\ave c$ \cite{RefWorks:127}.
Figure \ref{fig:GC_and_suskis_constDegHigh} shows link percolation results for networks for $\delta \in [0,1]$.  
For small values of  $\delta$ it appears that there is no community structure compatible with the weak links hypothesis, 
as 
the giant component is destroyed at the same point for both removal orders and $\tilde s$ remains very small. 
However, when $\delta \gtrsim 0.1$ the networks start to break faster when weak links are removed first and $\tilde s$ developes a finite signature of divergence, indicating that the weak links serve as bridges connecting communities. This
is also corroborated by the the rapid decrease in $\ave c$ when strong links are removed first (see inset). 
This effect was also verified similarly to Refs.~\cite{RefWorks:129,RefWorks:130} using the topological overlap measure,
which was found to increase clearly as a function of link weight when $\delta \gtrsim 0.1$ (not shown).

\begin{figure}[!tb]
\includegraphics[width=1.0\linewidth]{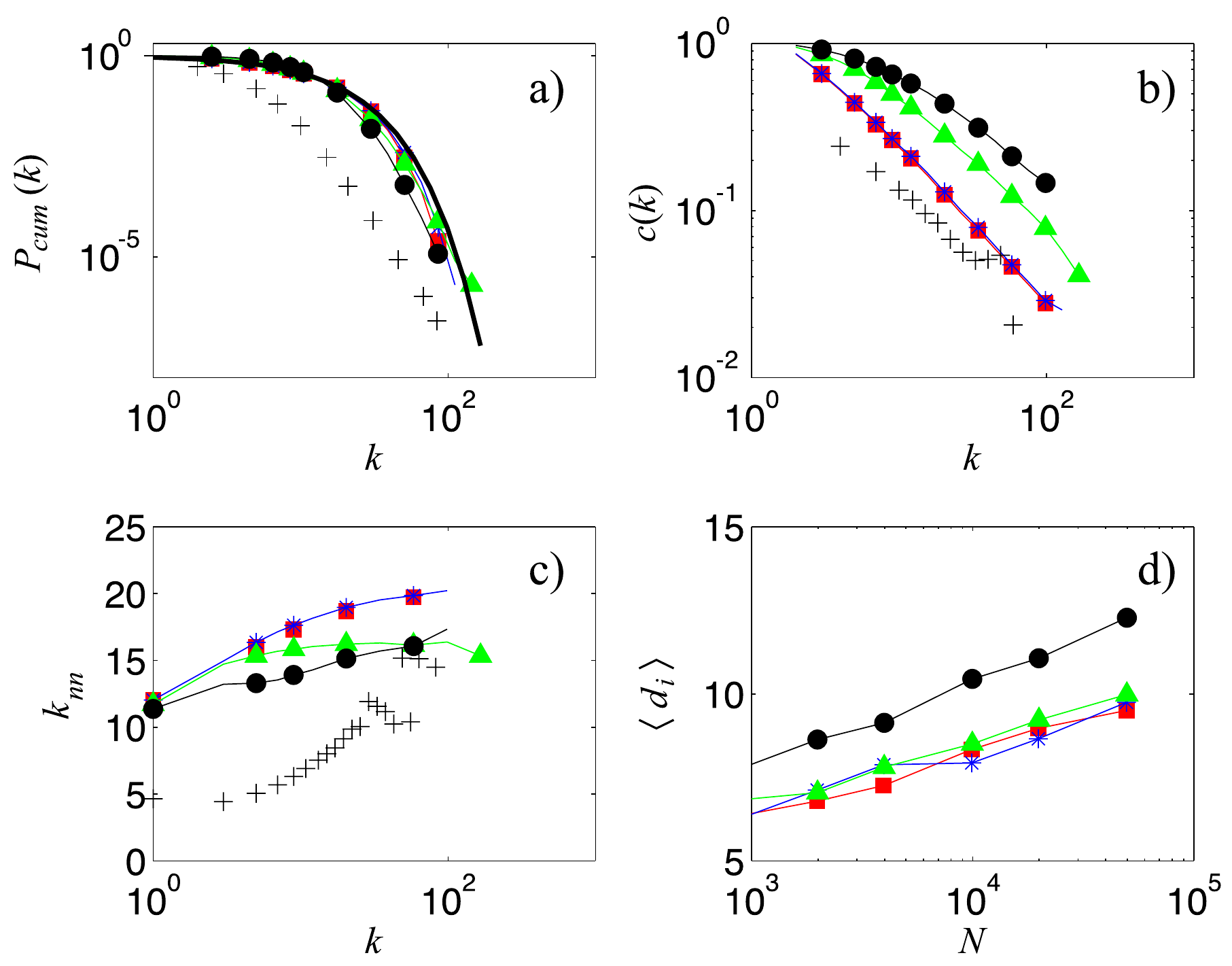}
\caption{(Color online) The model meets criteria ($i$)-($iv$) for social networks (see text) 
when $\delta \in [0, 1]$. (a) Degree distribution, (b) clustering, (c) average nearest neighbor degree $k_{nn}$, and (d) sampled
network diameter as a function of network size $N$.
Results for (a), (b), and (c) are averaged over 10 realizations of $N=5\times 10^4$  networks.
The corresponding distributions for the mobile phone call network  studied in Ref.~\cite{RefWorks:129}  are marked by + 
(note that this network has average degree $\ave k=3.14$),
the exponential fit in (a) is shown as a solid line, other symbols are as in Fig. \ref{fig:GC_and_suskis_constDegHigh}. }
\label{fig:jakaumia}
\end{figure}

Finally, as our model is inspired by mechanisms of social network formation, 
we investigate whether it reproduces also other properties of social networks. 
The picture emerging from analysis of a number of large data bases
\cite{RefWorks:129,RefWorks:130,RefWorks:132,RefWorks:137} shows several common features: $i$) degree distributions are skewed, $ii$) high-degree nodes are often connected to other high-degree nodes (assortative
mixing), $iii$) the average clustering coefficient $\langle c \rangle$ is high compared to random networks,  and
$iv$) the average shortest path lengths 
are small (the small-world property).
Figure \ref{fig:jakaumia} presents the basic distributions for network quantities with $\delta\in [0,1]$. 
The degree distribution was found to be almost independent of $\delta$. It has a quite fast decaying tail with 
the maximum degree $\sim 10^2$ for networks with 
$\ave k \approx 10$ and can be fitted reasonably with an exponential distribution. 
The networks show high clustering behaving almost as $c(k)\sim 1/k$. 
The model networks are also found to be assortative and the diameter grows as $\log N$, Fig. \ref{fig:jakaumia}(c,d).

In this paper we have introduced a simple weighted network model in which the weights are an essential part of the microscopic mechanisms. 
They establish a coupling between network structure and interaction strengths: addition of a new link depends on the existing weights, and once a 
new link is added the weights that led to its formation are strengthened. Communities will emerge only if this strenghtening is large enough, i.e., if nodes favor sufficiently their
 strong connections in the process of establishing new ones. 
Our study support the notion that communities result from initial structural fluctuations, which become amplified by repeated application of the microscopic processes. 
In addition to fulfilling the standard topological properties of social networks, the model networks exhibit a coupling between network topology and interaction strengths, which is compatible with the weak links hypothesis.  
Most importantly, this model provides a starting point for understanding the formation of communities in weighted networks.

\textbf{Acknowledgements:} J.M.K., J.-P.O., J.S. and K.K. acknowledge the Academy of Finland, the Finnish Center of Excellence program 2006-2011, proj. 213470.
J.-P.O. is supported by a Wolfson College, University of Oxford, Junior Research Fellowship. 
 J.K. was partially
supported by OTKA T049238 grant. 

\end{document}